\documentclass[10pt,twocolumn,letterpaper]{article}
\usepackage{cite}
\usepackage{amsmath,amssymb,amsfonts}
\usepackage{algorithmic}
\usepackage{graphicx}
\usepackage{textcomp}
\usepackage{xcolor}
\def\BibTeX{{\rm B\kern-.05em{\sc i\kern-.025em b}\kern-.08em
    T\kern-.1667em\lower.7ex\hbox{E}\kern-.125emX}}
\begin{document}

\title{Curvilinear Aperture Monopulse}

\author{Mark Story\\
\textit{Cyber Resilience and Intelligence Division} \\
\textit{Oak Ridge National Laboratory}\\
Oak Ridge, Tennessee, USA \\
storyma@ornl.gov}

\maketitle

\begin{abstract}
  By a symmetry argument, a synethic aperture radar collection along a
  linear path does not collect three-dimensional information about the
  scene.  However, it is known that vertical curvature can be used to
  derive some vertical position information.  This paper approaches
  the problem from a monopulse perspective, resulting in a
  non-iterative computation that commutes with efficient image
  formation algorithms. \footnote{Research sponsored by Oak Ridge
  National Laboratory, managed by UT-Battelle, LLC, for the
  U. S. Department of Energy.}
\end{abstract}


\section{Introduction}

Synthetic Aperture Radar (SAR) depicts radar reflectivity as a
two-dimensional image in range and azimuth relative to the aperture
traced out by the path of the platform.  A three-dimensional point
cloud can be created from two or more appropriately matched SAR
images, using interferometric SAR or radargrammetric stereo (as in,
for example, \cite{jakowatz_text}).  But for a SAR image from a linear
aperture, no information is present about the position of a scatterer
along a the circle centered on a line through that aperture and lying
on a plane perindicular to it.  This follows from simple symmetry
arguments.

However, when the aperture is not linear, the symmetry is broken.
Formation of a sharply-focused two-dimensional image requires
compensating for out-of-plane motion using assumed height of
scatterers, typically on the ground plane \cite{jakowatz_text}.
Conversely, objects in an image with height differing from the assumed
ground plane may appear to be out of focus.

This depth of focus issue for curvilinear apertures, together with its
implication for height estimation, has been most recently studied by
\cite{Doerry2021_haf}, \cite{Doerry2022_haf}.  The use of quadratic
phase errors for this purpose appears in \cite{Barnes_patent}, for a
particular maneuver.  More recent approaches include ``CLEAN''
\cite{Knaell1994}, \cite{Knaell1995_Tomo}, \cite{Knaell1995_Prac};
Basis Pursuit Denoising \cite{Austin2008}; and $\ell_1$-regularized
sparse reconstruction followed by an iterative model and subtract
(IMAS) step \cite{Jackson2012}.  These use computationally-intensive
iterative algorithms that rely on strong implicit or explicit
assumptions that the radar reflections are caused by a few discrete or
canonical scatterers.  In \cite{Andre2010}, 3D backprojection imaging
is applied to discrete scatterers.

In this paper, we instead approach a curvilinear aperture as a
phased-array monopulse antenna, using techniques related to
\cite{Henderson1985} and \cite{Henderson2005}.  We derive the
curvilinear aperture monopulse (CLAM) computations for vertical offset
in Section \ref{derivation}, under the assumption that there is a
single scatterer near a focus point, we compute its offset in the
vertical dimension.  The technique commutes with efficient image
formation algorithms as monopulse commutes with beamforming.  Section
\ref{examples} demonstrates the computation for an aperture with a
simple third-order polynomial vertical component and a monotonic
horizontal component.  A restriction analogous to focus arises in the
context of interference in Section \ref{interference}.  Section
\ref{conclusion} summarizes conclusions.

\section{Derivation}\label{derivation}

This section develops CLAM equations for a curvilinear aperture,
in a single-frequency setting.  Of course, practical radar pulses have
finite bandwidth to permit range discrimination.  But many radar
applications are sufficiently narrowband that within a compressed
pulse envelope (perhaps with time-delay focusing), the
single-frequency approximation is valid.

\subsection{Single frequency signal model}\label{signalmodel}

A receiver moves along an aperture defined by
\begin{equation}
  \begin{aligned}
    x &= x_0 + x_{\tau} + \Delta x \\
    y &= y_0 + \Delta y \\
    z &= z_0 + z_{\tau} + \Delta z,
  \end{aligned}
  \label{xyz}
\end{equation}
where $x_0, y_0, z_0$ are known fixed offsets, and where $x_\tau,
z_\tau$ describe the aperture as $\tau$ varies.  We require that
$x_\tau$ and $z_\tau$ have three derivatives.  $\Delta x, \Delta y,
\Delta z$ are unknown fixed offsets that we intend to measure, and we
are particularly interested in $\Delta z$.  (Of course, the radar
directly measures $\Delta y$, but $\Delta y$ is needed in the
computation for reasons that appear in Section \ref{parabola}.)
Nonlinear $z_\tau$ specifies a curvilinear aperture.  Nonlinear
$x_\tau$ does not significantly complicate the following derivation,
but our main interest is on the case of monotonic or linear $x_\tau$.
To simplify this derivation, $y$ does not vary with $\tau$.

Along this aperture, we measure a single-frequency radiating scalar
field given by
\begin{equation}
  E = A e^{jk(\frac{1}{p}ct - R_{\tau})},
  \label{wave}
\end{equation}
where $R_{\tau} = \sqrt{x^2 + y^2 + z^2}$ is the range from $x = y =
z$ to the origin.  Here, $E$ is the horizontal or vertical
polarization of the electric or magnetic field, $t$ is ``fast time'',
and $\tau$ is ``slow time''.  Coupling between $t$ and $\tau$ is
ignored, and we assume the platform does not move during the travel
time of the radar pulse.  Choose $p=1$ if the transmitter is
stationary or $p=2$ if the transmitter moves along the aperture with
the receiver.  

For typical SAR applications, the variation in $x_\tau$ and $z_\tau$
is large enough relative to $y$ that the plane wave approximation is
not valid.  However, the first order taylor approximation $\sqrt{1 +
  u} \approx 1 + \frac{1}{2}u$ is often sufficient, leading to
\begin{equation}
  2 y  R_\tau \approx 2 y^2 + x^2 + z^2.
  \label{parabola_range_alt}
\end{equation}
From here forward, we accept (\ref{parabola_range_alt}) as sufficient
and drop the approximation symbol.

\subsection{System of equations}

From (\ref{wave}), the derivative of E with respect to $\tau$ is
\begin{equation}
  E' = -jkR'_\tau E.
  \label{waveprime}
\end{equation}
(The ``prime'' symbol will be used throughout do denote
differentiation with respect to $\tau$.)  Taking the derivative of
(\ref{parabola_range_alt}) with respect to $\tau$ and expanding using
(\ref{xyz}),
\begin{equation}
  (y_0 + \Delta y)  R'_\tau = x'_\tau (x_0 + x_\tau + \Delta x) +
  z'_\tau (z_0 + z_\tau + \Delta z).
  \label{Rprime}
\end{equation}
Define $Q_\tau$ as the value of $R_\tau$ if the unknown $\Delta x,
\Delta y, \Delta z$ are set to zero, and then
\begin{equation}
  y_0 Q'_\tau = x'_\tau (x_0 + x_\tau) + z'_\tau (z_0 + z_\tau).
  \label{Qprime}
\end{equation}
Substituting (\ref{Qprime}) into (\ref{Rprime}),
\begin{equation}
  (y_0 + \Delta y) R'_\tau = x'_\tau \Delta x +
  z'_\tau \Delta z + y_0 Q_\tau',
  \label{RpmQp}
\end{equation}
and with (\ref{waveprime}),
\begin{equation}
  (y_0 + \Delta y) E' = -jkE (x'_\tau \Delta x + z'_\tau \Delta z + y_0 Q'_\tau),
  \label{Eprime}
\end{equation}
which is linear in $\Delta x$, $\Delta y$, and $\Delta z$:
\begin{equation}
  \begin{aligned}
    y_0 (E' + jk Q_\tau'E ) =& \\
    -jkE x'_\tau \Delta x &- E' \Delta y - jkE z'_\tau \Delta z.
  \end{aligned}
  \label{Eprime}
\end{equation}
Differentiating (\ref{Eprime}),
\begin{equation}
  \begin{aligned}
    y_0 (E'' + jk (Q_\tau'E)' ) =& \\
    -jk (E x'_\tau)' \Delta x & - E'' \Delta y
    -jk (E z'_\tau)' \Delta z ,
  \end{aligned}
  \label{Edoubleprime}
\end{equation}
and differentiating (\ref{Edoubleprime}),
\begin{equation}
  \begin{aligned}
    y_0 (E''' + jk (Q_\tau'E)'' ) =& \\
    -jk (E x'_\tau)'' \Delta x &- E''' \Delta y
    -jk (E z'_\tau)'' \Delta z.
  \end{aligned}
  \label{Etripleprime}
\end{equation}
Now (\ref{Eprime}), (\ref{Edoubleprime}), and (\ref{Etripleprime}) are
a system of three linear equations in $\Delta x$, $\Delta y$, and
$\Delta z$.  If we know the aperture, then $y_0$, as well as $x_\tau$,
$z_\tau$, $Q_\tau$, and their derivatives are known.  It remains to
compute or measure $E'$, $E''$, and $E'''$.

\subsection{Scalar field derivatives}

Direct measurement of the first three derivatives of the scalar field
is unreasonable in many settings.  But we can estimate the derivative
across the aperture as follows.

Let $w$ be a finite-length discrete window function with support over
the slow-time aperture.  Let $h$ be the single-frequency
backprojection function for a hypothesized scatterer at
$(x_0, y_0, z_0)$:
\begin{equation}
  h(\tau) = e^{-jkQ_\tau}.
  \label{nbbp}
\end{equation}
In the azimuthal direction, backprojection image formation of a pixel
at $(x_0, y_0, z_0)$, with a window, is given by the integral
\begin{equation*}
  \int_{-T}^T w(\tau) h(\tau) E d\tau,
\end{equation*}
where $\pm T$ are the values of $\tau$ specifying the beginning and end
of the aperture.

For convenience, we will write $h_w(\tau) = w(\tau) h(\tau)$.  Using
integration by parts, if $w(\tau)$ is zero except on $[-T,T]$, and if
its derivative exists everywhere, then
\begin{equation}
  \begin{aligned}
  \int_{-T}^T h_w(\tau) E' d\tau &= -\int_{-T}^T h'_w(\tau) E d\tau \\
  \int_{-T}^T h_w(\tau) E'' d\tau &= +\int_{-T}^T h''_w(\tau) E d\tau \\
  \int_{-T}^T h_w(\tau) E''' d\tau &= -\int_{-T}^T h'''_w(\tau) E d\tau.
  \end{aligned}
  \label{intEprimes}
\end{equation}
Following \cite{Henderson2005}, if $w$ is chosen to have the form
\begin{equation*}
  w(\tau) = w_\text{base}(\tau) * (\delta_{-\frac{1}{2}s} + \delta_{\frac{1}{2}s})
\end{equation*}
(where $\delta_u$ is the dirac delta shifted to $u$), then for small
$s$, its derivative is approximated by
\begin{equation*}
  w_1(\tau) = w_\text{base}(\tau) * \frac{1}{s} (\delta_{-\frac{1}{2}s} - \delta_{\frac{1}{2}s}).
\end{equation*}
Applying this approximation multiple times, if
\begin{equation}
  \begin{aligned}
    w_0(\tau) &= w_\text{base}(\tau) * (\delta_{-\frac{3}{2}s}
    + 3 \delta_{-\frac{1}{2}s} +
    3 \delta_{\frac{1}{2}s} + \delta_{\frac{3}{2}s}) \\
    w_1(\tau) &= w_\text{base}(\tau) * \frac{1}{s} (\delta_{-\frac{3}{2}s}
    + \delta_{-\frac{1}{2}s} -
     \delta_{\frac{1}{2}s} - \delta_{\frac{3}{2}s}) \\
     w_2(\tau) &= w_\text{base}(\tau) * \frac{1}{s^2} (\delta_{-\frac{3}{2}s}
     - \delta_{-\frac{1}{2}s} -
     \delta_{\frac{1}{2}s} + \delta_{\frac{3}{2}s}) \\
     w_3(\tau) &= w_\text{base}(\tau) * \frac{1}{s^3} (\delta_{-\frac{3}{2}s}
     - 3 \delta_{-\frac{1}{2}s} +
    3 \delta_{\frac{1}{2}s} - \delta_{\frac{3}{2}s})
  \end{aligned}
  \label{wprimes}
\end{equation}
(where $*$ denotes convolution), then the $i^{\text{th}}$ derivative
of $w$ is approximated by $w_i$ for small $s$, for $i = 0, 1, 2, 3$.
Turning attention to derivatives of the narrowband backprojection
function,
\begin{equation}
  \begin{aligned}
    h'(\tau) &= -jk Q_\tau' h(\tau) \\
    h''(\tau) &= ((jk Q_\tau')^2 - jk Q''_\tau) h(\tau) \\
    h'''(\tau) &= ( -(jk Q_\tau')^3 + 3 (jk)^2 Q'_\tau Q''_\tau - jk Q'''_\tau) h(\tau)
  \end{aligned}
\end{equation}
and inserting the approximations in (\ref{wprimes}),
\begin{equation}
  \begin{aligned}
    h'_w(\tau) &\approx w_1(\tau) h(\tau) + w_0(\tau) h'(\tau)\\
    h''_w(\tau) &\approx w_2(\tau) h(\tau) + 2w_1(\tau) h'(\tau)
    + w_0(\tau) h''(\tau)\\
    h'''_w(\tau) &\approx w_3(\tau) h(\tau) + 3w_2(\tau) h'(\tau)\\
    &\qquad + 3w_1(\tau) h''(\tau) + w_0(\tau) h'''(\tau).
  \end{aligned}
\end{equation}
We can then compute integrals of each side of (\ref{Eprime}),
(\ref{Edoubleprime}), and (\ref{Etripleprime}), and
\begin{equation}
  \mathbf{M} \Delta \mathbf{r} =
  \begin{bmatrix}
      M_{00} & M_{01} & M_{02} \\
      M_{10} & M_{11} & M_{12} \\
      M_{20} & M_{21} & M_{22}
  \end{bmatrix}
  \begin{bmatrix}
    \Delta x \\
    \Delta y \\
    \Delta z
  \end{bmatrix}
  =
  \begin{bmatrix}
    b_{0} \\
    b_{1} \\
    b_{2}
  \end{bmatrix}
  = \mathbf{b},
  \label{maineq}
\end{equation}
where
\begin{equation*}
    M_{00} = -jk \int_{-T}^T h_w(\tau) x'_\tau E d\tau,
\end{equation*}
\begin{equation*}
    M_{01} = -\int_{-T}^T h'_w(\tau) E d\tau ,
\end{equation*}
\begin{equation*}
    M_{02} = -jk \int_{-T}^T h_w(\tau) z'_\tau E d\tau ,
\end{equation*}
\begin{equation*}
    M_{10} = -jk \int_{-T}^T h'_w(\tau)) x'_\tau E d\tau ,
\end{equation*}
\begin{equation*}
    M_{11} = -\int_{-T}^T h''_w(\tau) E d\tau ,
\end{equation*}
\begin{equation*}
    M_{12} = -jk \int_{-T}^T h'_w(\tau)) z'_\tau E d\tau ,
\end{equation*}
\begin{equation*}
    M_{20} = -jk \int_{-T}^T h''_w(\tau) x'_\tau E d\tau ,
\end{equation*}
\begin{equation*}
    M_{21} = -\int_{-T}^T h'''_w(\tau) E d\tau ,
\end{equation*}
\begin{equation*}
    M_{22} = -jk \int_{-T}^T h''_w(\tau) z'_\tau E d\tau ,
\end{equation*}
\begin{equation*}
    b_{0} = y_0 \int_{-T}^T (h'_w(\tau))' + jk h_w(\tau) Q'_\tau) E d\tau ,
\end{equation*}
\begin{equation*}
    b_{1} = y_0 \int_{-T}^T (h''_w(\tau))'' + jk h'_w(\tau) Q'_\tau) E d\tau ,
\end{equation*}
\begin{equation*}
    b_{2} = y_0 \int_{-T}^T (h'''_w(\tau) + jk h''_w(\tau) Q'_\tau) E d\tau.
\end{equation*}


For apertures critically sampled uniformly in $\tau$ at a sample rate
of $s$, the integrals above can be computed as sums.  When
$\mathbf{M}$ is invertible, $\Delta \mathbf{r} = \mathbf{M^{-1} b}$.
Our objective, $\Delta z$, is the third element of $\Delta
\mathbf{r}$.

\subsection{Computation using efficient image formation algorithms}

The integral in each $M_{ij}$ term and each $b_{i}$ term in
(\ref{maineq}), together with $h$, is the azimuth portion of a
backprojection image computation, for a single pixel, for the scalar
field $E$, under some window function.  For example, for $M_{00}$, the
window is $-jk w_0(\tau) x'_\tau$.  An efficient image formation
technique like Polar Format Algorithm commutes with the operations
above, and is a good approximation for backprojection.  So in
principle, $\mathbf{M}$ and $\mathbf{b}$ can be computed for all
pixels in a scene, by performing an image formation operation for each
$M_{ij}$ and each $b_{i}$.  We will see in Section \ref{interference}
that interference can corrupt pixels that are not notably brighter
than their neighbors.

For sufficiently large images, the approximation of
(\ref{parabola_range_alt}) may become invalid for large $x$ or $z$.
In this case, it may be necessary to vary $x_0$ and $z_0$ to keep the
approximation valid.  Handling this situation, while leveraging an
efficient image formation technique, is beyond the scope of this
paper.

\section{Examples}\label{examples}

\subsection{Asymmetric third-order polynomial aperture}

\begin{figure}[tbp]
\centerline{\includegraphics[width=.9\linewidth]{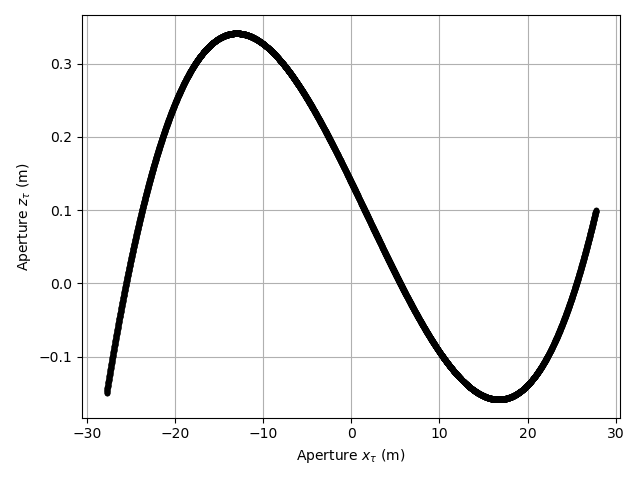}}
\caption{A third-order polynomial aperture of length 55.5m and height
  0.5m; with zeros at -25.54m, 5.55m, and 25.54m.}
\label{cubicap}
\end{figure}

\begin{figure}[tbp]
\centerline{\includegraphics[width=.9\linewidth]{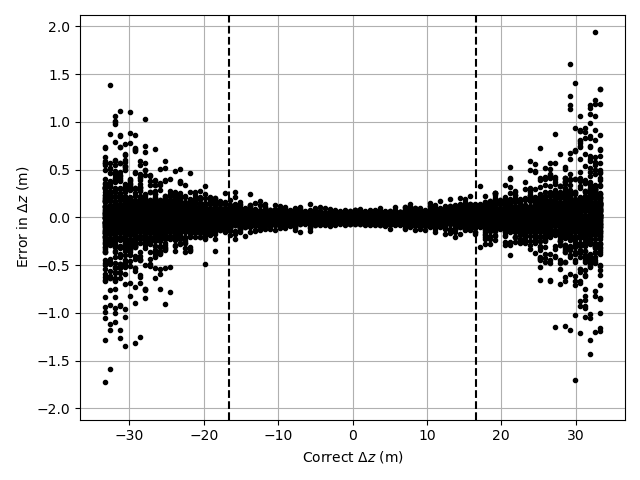}}
\caption{Error in computed $\Delta z$, varying $\Delta x$ and $\Delta
  y$, using (\ref{maineq}).  The computation is accurate within the
  vertical resolution of a circumscribing rectangle (marked by dashed
  lines), and not unreasonably subject to small amounts of noise or to
  variations in $\Delta x$ and $\Delta y$.}
\label{cubicerr}
\end{figure}

Consider the third-order polynomial aperture pictured in Fig.
\ref{cubicap}.  Its length is 55.5m, and its height is 0.5m.  Assume a
center frequency of 9 GHz, and a range of 1km.  A linear aperture of
this length would have a horizontal resolution of 0.3m.  Choose a hann
window for $w_\text{base}(\tau)$.  Using a single-frequency point-scatterer
simulation, setting $\Delta x = \Delta y = 0$, the computed vertical
position matches the correct position with $\Delta z$ out to and
somewhat beyond the dashed lines at $\pm 16.7m$, which are the edges
of vertical resolution of a hypothetical rectangular aperture
circumscribing the curve.  Noise was added to each time sample, at an
amplitude of 10\% of that of the signal.  Varying $\Delta x$ over the
horizontal resolution of a linear aperture of this length ($\pm
0.15\text{m}$), and varying $\Delta y$ over the same amount (as might
be expected for a SAR system), the error in computed $\Delta z$ is
shown in Fig. \ref{cubicerr}.  So the approximations we have
introduced are capable of measuring $\Delta z$ to a high degree of
accuracy in this setting, varying $\Delta x, \Delta y$ over the space
of a SAR pixel, for a single scatterer, with a small amount of noise.

\begin{figure}[tbp]
\centerline{\includegraphics[width=.9\linewidth]{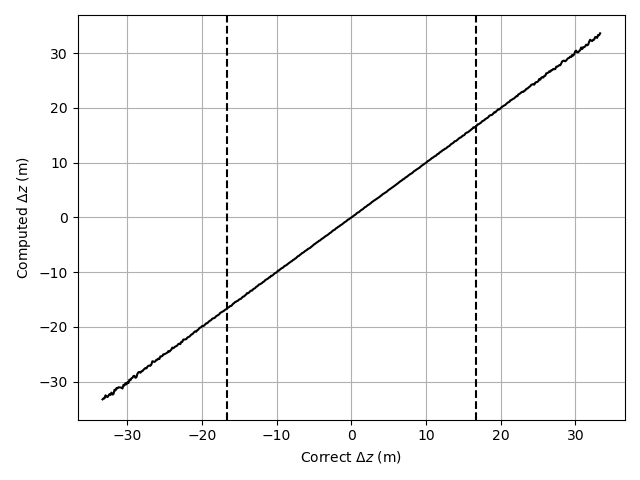}}
\caption{Computed and correct $\Delta z$.  The computation is
  accurate, and is not unreasonably subject to small amounts of
  noise.}
\label{cubicout}
\end{figure}

\subsection{Parabolas aperture ambiguities}\label{parabola}

For a symmetric second-order polynomial aperture, either a height
offset or a small range offset causes similar wavefronts to reach the
aperture.  For the full computation of (\ref{maineq}), the ambiguity
causes an ill-conditioned matrix.  But if we make the assumption that
$\Delta y$ is zero, we can modify (\ref{maineq}) to remove both
$\Delta y$ and (\ref{Etripleprime}), leaving
\begin{equation}
  \begin{bmatrix}
      M_{00} & M_{02} \\
      M_{10} & M_{12}
  \end{bmatrix}
  \begin{bmatrix}
    \Delta x \\
    \Delta z
  \end{bmatrix}
  =
  \begin{bmatrix}
    b_{0} \\
    b_{1}
    \end{bmatrix}.
  \label{modeq}
\end{equation}

\begin{figure}[tbp]
\centerline{\includegraphics[width=.9\linewidth]{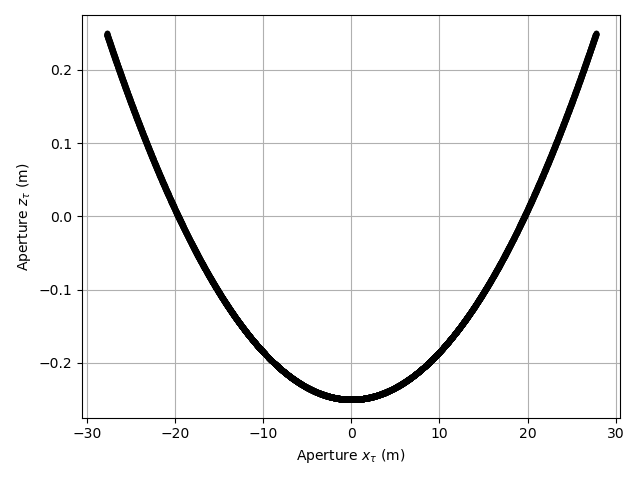}}
\caption{A second-order polynomial aperture of length 55.5m and height
  0.5m; with zeros at $\pm$ 19.63m.}
\label{parabap}
\end{figure}

\begin{figure}[tbp]
\centerline{\includegraphics[width=.9\linewidth]{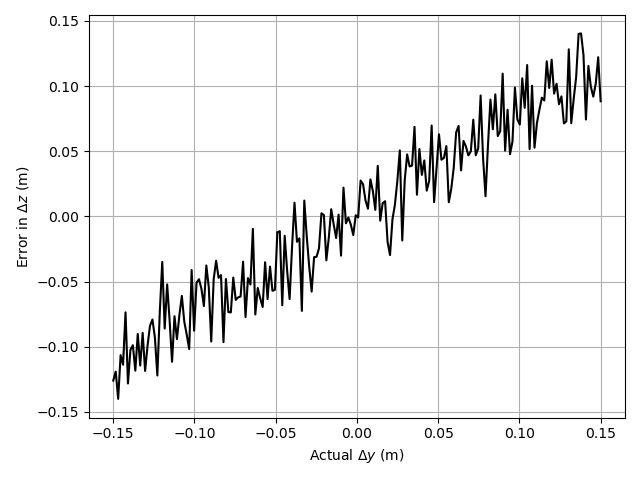}}
\caption{Error in computed $\Delta z$, for the second-order polynomial
  aperture of Fig. \ref{parabap}, caused by small offsets in actual
  $\Delta y$, using (\ref{modeq}).  Small errors due to noise are also
  present.}
\label{prng}
\end{figure}

\begin{figure}[tbp]
\centerline{\includegraphics[width=.9\linewidth]{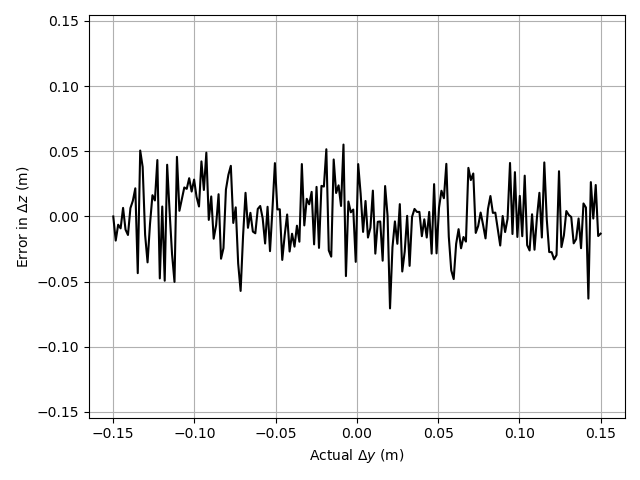}}
\caption{Error in computed $\Delta z$, for the third-order polynomial
  aperture of Fig. \ref{cubicap}, using (\ref{maineq}).
  Small errors are present due to noise, but offsets in actual $\Delta
  y$ do not correlate with errors in measured $\Delta z$.}
\label{crng}
\end{figure}

\begin{figure}[tbp]
\centerline{\includegraphics[width=.9\linewidth]{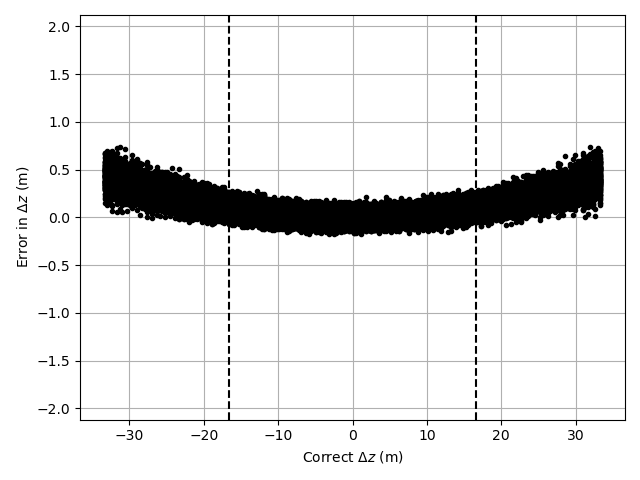}}
\caption{Error in computed $\Delta z$, for the second-order polynomial
  aperture of Fig. \ref{parabap}, using (\ref{modeq}), varying
  $\Delta x$ and $\Delta y$.  The computation is accurate within the
  vertical resolution of a circumscribing rectangle, and not
  unreasonably subject to small amounts of noise.  But some systematic
  error is present.}
\label{paraberr}
\end{figure}

A symmetric second-order polynomial aperture is shown in Fig.
\ref{parabap}.  The aperture height and length are the same as that of
Fig. \ref{cubicap}.  Using (\ref{modeq}), measured $\Delta z$ varies
with small range offsets, as shown in Fig. (\ref{prng}).  So the
impact of dropping $\Delta y$ is that the vertical position estimate
varies with small range offsets.  The full CLAM computation using the
aperture of Fig. (\ref{cubicap}) is immune to this ambiguity, as
shown in (\ref{crng}).  In certain applications and for certain
apertures, the advantage of using a parabolic aperture may outweigh
the impact of this ambiguity.  In the case pictured, the vertical
error is comparable to the size of the azimuthal resolution, which is
typically comparable to the azimuthal resolution.  But for some
applications, this may be acceptible.  Larger vertical apertures yield
smaller ambiguities, for the same horizontal aperture.  See Fig.
(\ref{paraberr}) for results varying $\Delta x, \Delta y$ as in Fig.
(\ref{cubicerr}) The size of the errors are comparable for these
parameters.

\subsection{Interference and glint}\label{interference}

The computation for $\Delta x, \Delta y, \Delta z$ in (\ref{maineq})
is nonlinear in the scalar field $E$, because $E$ is present in all
terms of $\mathbf{M}$ and $\mathbf{b}$ in (\ref{maineq}).  For this
reason, if multiple scatterers contribute to $E$, the computed values
of $\Delta x, \Delta y, \Delta z$ are not a linear combination of the
values that would be computed in the single-scatterer case.  Nor are
they guaranteed to be between or even near the correct values.  This
phenomenon is known as ``glint'' in the context of monopulse
\cite{Henderson2005}.

\begin{figure}[tbp]
\centerline{\includegraphics[width=.9\linewidth]{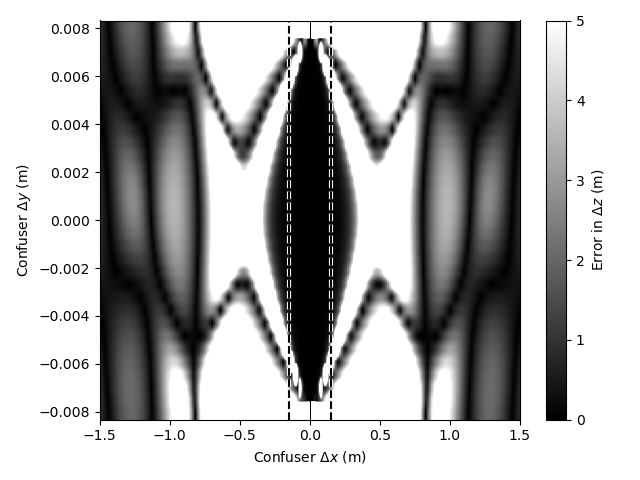}}
\caption{Error in computed $\Delta z$ due to confuser scatterers.
  Both the central scatterer and the confuser scatterer are at $\Delta
  z = 0$, but the confuser scatterer position varies over half the
  wavelength.  The dashed line indicates the horizontal resolution of
  the aperture.  The confuser scatterer can cause large errors.}
\label{glint}
\end{figure}

Consider a scatterer at $\Delta x = 0, \Delta y = 0, \Delta z = 0$.
If a ``confuser'' scatter is placed at the same $\Delta z = 0$, but at
some other $\Delta x, \Delta y$, then the measured $\Delta z$ is not
necessarily near $0$.  For the aperture of Fig. (\ref{cubicap}), the
size of the error is shown in Fig. \ref{glint}.  $\Delta y$ is
varied over half of a wavelength, because the phenomenon appears to be
caused by the phase difference of the confuser.  The horizontal
resolution width is marked by dashed lines.  The error is small for
most phases, for most confuser scatterers within the resolution width.
So a confuser scatterer in the same pixel is likely to result in the
correct measurement.  For confuser scatterers several resolution
widths away, the directional effects of the backprojection function
$h(\tau)$ rejects the energy from the confuser, and the measurement is
approximately correct.  However, for confuser scatterers only a few
resolution widths away, the error can be quite large.  This is because
the beampatterns of many of the terms in (\ref{maineq}) are wider than
a resolution cell.  Information from adjacent resolution cells are
needed to make the measurement correctly.  As a result of this, these
computations will only be valid for scatterers that are particularly
bright relative to their immediate surroundings in azimuth.  SAR
images often have pixels that are notably brighter than their
neighbors, and this assumption is the basis of determining whether an
image is in focus.

\begin{figure}[tbp]
\centerline{\includegraphics[width=.9\linewidth]{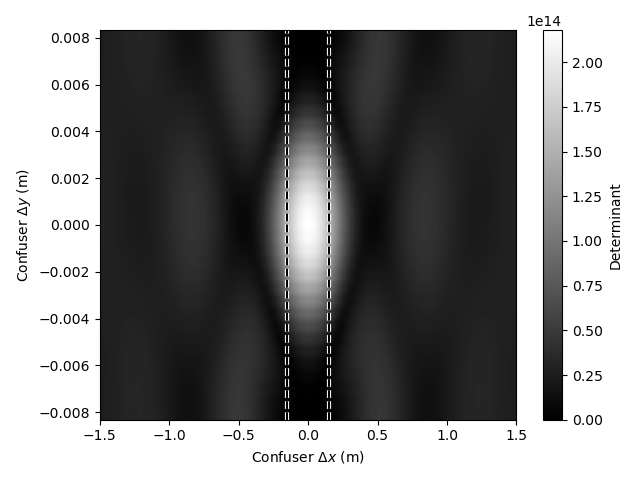}}
\caption{Determinant of $\mathbf{M}$ for the same case as Fig.
  \ref{glint}.  Where the determinant in this figure is large, the
  error in Fig. \ref{glint} is small.}
\label{glintdet}
\end{figure}

In a sensing context, it may be difficult to guarantee that only a
single scatterer is represented in a location.  We need a method to
detect when a measurement is corrupted by ``glint'' from other
scatterers.  The determinant from the left side of (\ref{maineq}) is
shown in Fig. \ref{glintdet}.  Where this determinant is large, the
glint error in Fig. \ref{glint} is small.  This suggests that small
determinants may be used to detect measurements corrupted by
interference from other scatterers.

\section{Conclusion}\label{conclusion}

Using computations given in this paper, with a curvilinear aperture,
one can compute elevation of a single scatterer, provided the matrix
$\mathbf{M}$ from (\ref{maineq}) is invertible.  The monopulse-based
computations are non-iterative, and commute with efficient image
formation algorithms.  A nearby confuser scatterer of similar
amplitude can corrupt the measurement, so elevation will only be
available for pixels that are notably brighter than their neighbors.
This is analogous to the focus assumption in other work on curvilinear
apertures, in that the concept of ``focus'' implies that for the
correct image, certain pixels are notably brighter than their
neighbors.  For a pixel subject to this corruption, the determinant of
$\mathbf{M}$ is small, serving as a warning.


\end{document}